\documentclass[8.5pt,twoside,twocolumn]{article}
\usepackage[super,sort&compress,comma]{natbib} 
\usepackage{mhchem}
\usepackage{times,mathptmx}
\usepackage{sectsty}
\usepackage{balance} 

\usepackage{graphicx} %eps figures can be used instead
\usepackage{lastpage}
\usepackage[format=plain,justification=raggedright,singlelinecheck=false,font=small,labelfont=bf,labelsep=space]{caption} 
\usepackage{amsmath}
\usepackage{enumitem}
\usepackage{graphicx}
\usepackage{subfigure}
\usepackage{float}
\newcommand{\lb}{\lambda_\mathrm{B}}
\newcommand{\lbw}{\lambda_\mathrm{B}^\mathrm{w}}
\newcommand{\lbo}{\lambda_\mathrm{B}^\mathrm{o}}

\newcommand{\PhiD}{\phi_{\mathrm{D}}}

\begin{document}

\makeatletter 
\def\subsubsection{\@startsection{subsubsection}{3}{10pt}{-1.25ex plus -1ex minus -.1ex}{0ex plus 0ex}{\normalsize\bf}} 
\def\paragraph{\@startsection{paragraph}{4}{10pt}{-1.25ex plus -1ex minus -.1ex}{0ex plus 0ex}{\normalsize\textit}} 
\renewcommand\@biblabel[1]{#1}            
\renewcommand\@makefntext[1]% 
{\noindent\makebox[0pt][r]{\@thefnmark\,}#1}
\makeatother 
\renewcommand{\figurename}{\small{Fig.}~}
\sectionfont{\large}
\subsectionfont{\normalsize}

\twocolumn[
  \begin{@twocolumnfalse}
\noindent\LARGE{\textbf{Anomalous system-size dependence of electrolytic cells with an electrified  oil-water interface}}
\vspace{0.6cm}

\noindent\large{\textbf{Marise Westbroek,$^{\ast}$\textit{$^{a}$} Niels Boon,\textit{$^{a}$} and
Ren\'e van Roij\textit{$^{a}$}}}\vspace{0.5cm}
%Please note that \ast indicates the corresponding author(s) but no footnote text is required. 

%\noindent\textit{\small{\textbf{Received Xth XXXXXXXXXX 20XX, Accepted Xth XXXXXXXXX 20XX\newline
%First published on the web Xth XXXXXXXXXX 200X}}}

%\noindent \textbf{\small{DOI: 10.1039/b000000x}}
%\vspace{0.6cm}
%Please do not change this text.

\noindent \normalsize{Manipulation of the charge of the dielectric interface between two bulk liquids not only enables the adjustment of the interfacial tension but also controls the storage capacity of ions in the ionic double layers adjacent to each side of the interface. However, adjusting this interfacial charge by static external electric fields is difficult. If exchange of electrons between external electrodes and ionic solution, i.e. an electric current, is impossible, a static equilibrium is reached. The external electric fields are readily screened by ionic double layers in the vicinity of the external electrodes, leaving the liquid-liquid interface at macroscopic distances from the electrodes unaffected. In this study we show theoretically, in agreement with recent experiments, that control over this surface charge at the liquid-liquid interface is nonetheless possible for macroscopically large but finite \emph{closed} systems, even when the distance between the electrode and interface is orders of magnitude larger than the Debye screening lengths of the two liquids. We identify a crossover system-size  below which the interface and the electrodes are effectively coupled. Our calculations of the interfacial tension for various electrode potentials are in good agreement with recent experimental data. 
}
\vspace{0.5cm}
 \end{@twocolumnfalse}
  ]

\section{Introduction}
\footnotetext{\textit{$^{a}$~Institute for Theoretical Physics, Leuvenlaan 4, 3584 CE Utrecht, The Netherlands. Tel: +447881983074; E-mail: mjw13@imperial.ac.uk}}

The ion distribution in the vicinity of charged surfaces in a liquid electrolyte  is a classic and important topic within physical chemistry. This field goes back  to at least the 1910s when Gouy \cite{gouy} and Chapman \cite{chap} identified the existence of a diffuse ionic cloud in the vicinity of the charged surface. This ionic cloud with a net charge exactly opposite to that of the surface has a thickness (now called the Debye screening length) typically in the range of 1-1000 nm depending on the ion concentration and the dielectric constant of the electrolyte. This implies that the effect of a static external charge immersed in a bulk electrolyte is only noticeable at distances smaller than several Debye lengths; at larger distances the external charge is fully screened by its surrounding ionic cloud. Indeed, it is well known that the effective electrostatic interactions between colloidal particles stem from their overlapping ionic clouds, thereby setting the interaction range equal to the Debye length of the supporting electrolyte \cite{israel}.  The notion of ionic screening  also implies that an electrolyte in between two planar electrodes can (in the absence of chemical reactions) only be manipulated by a static applied voltage if the electrode-electrode separation is of the order of the Debye length or smaller; macroscopic electrode separations, e.g. on the centimetre scale much larger than any typical Debye length,  lead to two fully screened decoupled electrodes sandwiching a bulk electrolyte that is insensitive to the applied static voltage (time-dependent voltages in which ionic clouds need to be built up can have a much longer range). 

Building on the notion of ionic screening, one would at first sight also expect that a planar interface between two demixed bulk electrolytes (e.g. oil and water) sandwiched by two planar electrodes in the geometry of an electrolytic cell cannot be manipulated by the applied voltage if both electrodes are at a macroscopic distance from the interface. Recent experiments, however, challenge this expectation. It was shown that oil-water interfaces,  which  in the absence of any external potential exhibit two back-to-back ionic double layers due to a repartitioning of the ions \cite{zwanikken}, can actually be electrified by `remote'  external electrodes \cite{laanait2010, schloss2012}. In particular, it was shown that the oil-water surface tension could be modified by applying a voltage across electrodes separated from the interface by several centimetres while the Debye lengths are orders of magnitude smaller \cite{schloss2012}. In the present article we will provide a theoretical explanation of these observations by extending the classical Gouy-Chapman solution of the Poisson-Boltzmann equation to include two electrodes, an oil-water interface, and four ionic species each with their own affinity for oil and water as described by a Born self energy difference between an ion in oil and water. 

Before embarking on a detailed theoretical analysis, we first consider an extreme case that qualitatively illustrates the possible  emergence of a  large (macroscopic) length scale in this problem. Imagine a demixed oil-water system with two hydrophilic ion species that cannot penetrate into the oil and two hydrophobic ion species that cannot penetrate into the water. The impossibility of ion migration implies that both phases are constrained to be charge neutral, not only in bulk but even if they are put  in contact in a macroscopic electrolytic cell of the type cathode-water-oil-anode. So upon the application of a voltage between the cathode and the anode, the cathode will be screened by an excess of hydrophilic cations  and the anode by an excess of the hydrophobic anions. However, global neutrality of the individual volumes of water and oil causes the formation of a back-to-back double layer of ionic charge at the oil-water interface, with an excess of hydrophilic anions at the waterside of the interface and an excess of hydrophobic cations at oil side. The neutrality constraint imposes the magnitude of these oil-water ionic excess charges to be identical to that on the electrodes. In other words, in this limiting case the charge of the oil-water interface can be perfectly tuned by the applied voltage across the electrodes, even at macroscopic distances from the interface. By contrast, if at least one of the ionic species is `sufficiently' soluble in both oil and water, then the neutrality constraint only  applies to the oil-water system as a whole because ionic excess charge can migrate from one electrode to the other thereby leaving the oil-water interface unaffected (if the Debye lengths are much smaller than the cell-size). These two extreme cases show that a crossover from a microscopic to a macroscopic length scale is to be expected. We will show below that this length scale is of the order of  $(|\sigma|/\rho)\exp(|f|)$ with $\pm e\sigma $ the surface charge density of the electrodes, $\rho$ a typical salt concentration, and $|f|$ the magnitude of the smallest Born self energy of the four ionic species (in units of the thermal energy $k_BT$). Clearly $|\sigma|/\rho$ is a microscopic length scale, but with $|f|$ varying from order unity up to 20-30 the exponential dependence on $|f|$ gives rise to a huge regime of lengths that are strictly speaking microscopic but that can easily exceed any realistic macroscopic system size. In cases that this new length  scale exceeds the system size, the macroscopic system is anomalously `small'  such that `remote'  electrodes can modify the oil-water interface statically.

\section{Poisson-Boltzmann theory of an electrified oil-water interface}
\subsection{Gouy-Chapman theory for a single electrode}
Before considering the actual system of interest in this study, the electrified oil-water interface as illustrated in Fig. 1, we remind ourselves of the simpler problem of a single planar electrode in contact with a half-space of a 1:1 electrolyte, treated within Poisson-Boltzmann theory for point ions. Assuming lateral translational invariance, and denoting the perpendicular distance to the electrode by $z>0$, we wish to calculate the electrostatic potential $\Psi(z)$ and equilibrium concentration profiles of the  cations and anions $\rho_+(z)$ and $\rho_-(z)$, respectively. Setting the potential far from the electrode to zero, $\Psi(\infty)=0$, and denoting the bulk ion concentration by $\rho$, so that $\rho_+(\infty)=\rho_-(\infty)\equiv\rho$ by bulk neutrality, we relate the ion distributions and the electric potential for $z>0$ by the Boltzmann distribution $\rho_{\pm}(z)=\rho\exp[\mp \phi(z)]$ with the dimensionless potential $\phi(z)=e\Psi(z)/k_BT$. Here $e$ is the proton charge, $T$ the temperature, and $k_B$ the Boltzmann constant. The two Boltzmann distributions are complemented by the Poisson equation $\phi''(z)=-4\pi\lambda_B[\rho_+(z)-\rho_-(z)+\sigma\delta(z)]$ for $z\geq 0$, where a prime denotes a derivative with respect to $z$, where $\lambda_B=e^2/\epsilon k_BT$ is the Bjerrum length (in Gaussian units) of the solvent in terms of its relative dielectric constant $\epsilon$, and where the surface charge density in the plane $z=0$ of the electrode is given by $e\sigma$.
Combining these results gives the Poisson-Boltzmann equation with its boundary conditions
\begin{eqnarray}
\phi''(z)&=&\kappa^2\sinh\phi(z); \label{pb1}\\
\phi(\infty)&=&0;\nonumber\\
\phi'(0^+)&=&-4\pi\lambda_B\sigma,\nonumber
\end{eqnarray}
where the screening parameter is defined as $\kappa^2=8\pi\lambda_B\rho$. This closed set of equations can be solved analytically and yields \cite{gouy,chap,vanroij}
\begin{equation}\label{eq:pbsolution}
\phi(z)=  2\log\left[\frac{1+\gamma\exp(-\kappa z)}{1-\gamma\exp(-\kappa z)}\right],
\end{equation}
where the integration constant is given by
\begin{equation}\label{eq:gamma}
\gamma = \frac{\sqrt{(2\pi \lb \kappa^{-1} \sigma)^2+1}-1}{2\pi \lb \kappa^{-1} \sigma}.
\end{equation}
Note that $-1<\gamma<1$, that $\gamma\propto\sigma$ in the linear screening (low surface charge) regime $|\sigma|\ll \sigma^*\equiv (\pi\lambda_B\kappa^{-1})^{-1}$, and that $|\gamma|$ approaches unity in the strongly nonlinear screening limit (high surface charge) $|\sigma|\gg\sigma^*$.  The dimensionless potential  $\phi(z)$ of Eq.(\ref{eq:pbsolution}) and the associated ion distributions $\rho_{\pm}(z)=\rho\exp[\mp\phi(z)]$ describe the well-known diffuse ionic screening cloud of  typical thickness $\kappa^{-1}$ (the Debye length) in the vicinity of the electrode. In the present context we are interested in the cation and anion adsorptions, i.e. the excess number of ions per unit electrode area, defined by $\Gamma_\pm \equiv \int_0^{\infty} \mathrm{d}z~~(\rho_\pm(z) - \rho)$. It follows from Eq.(\ref{eq:pbsolution}) and (\ref{eq:gamma}) that
\begin{equation}\label{eq:Gamma}
\Gamma_\pm = \frac{\mp 4\rho}{\kappa}
\frac{\gamma}{1\pm\gamma}.
\end{equation}
One checks that $\Gamma_+ -\Gamma_-+\sigma =0$, such that charge neutrality is satisfied. We note, however, that the total ion adsorption $\Gamma_++\Gamma_-=\sigma^*/(\gamma^{-2}-1)$ depends nontrivially on the total surface charge: it is vanishingly small in the linear-screening regime but grows with  increasing surface charge to become of the same order as $\sigma^*$ if $\gamma\simeq 0.5$, and it diverges in the limit of highly charged surfaces (where the underlying assumption of point ions breaks down. This limit is not of concern in this study).

\begin{figure}[h!]
	\centering
\begin{subfigure}
	\centering
	\includegraphics[width=0.5\textwidth,height=40mm]{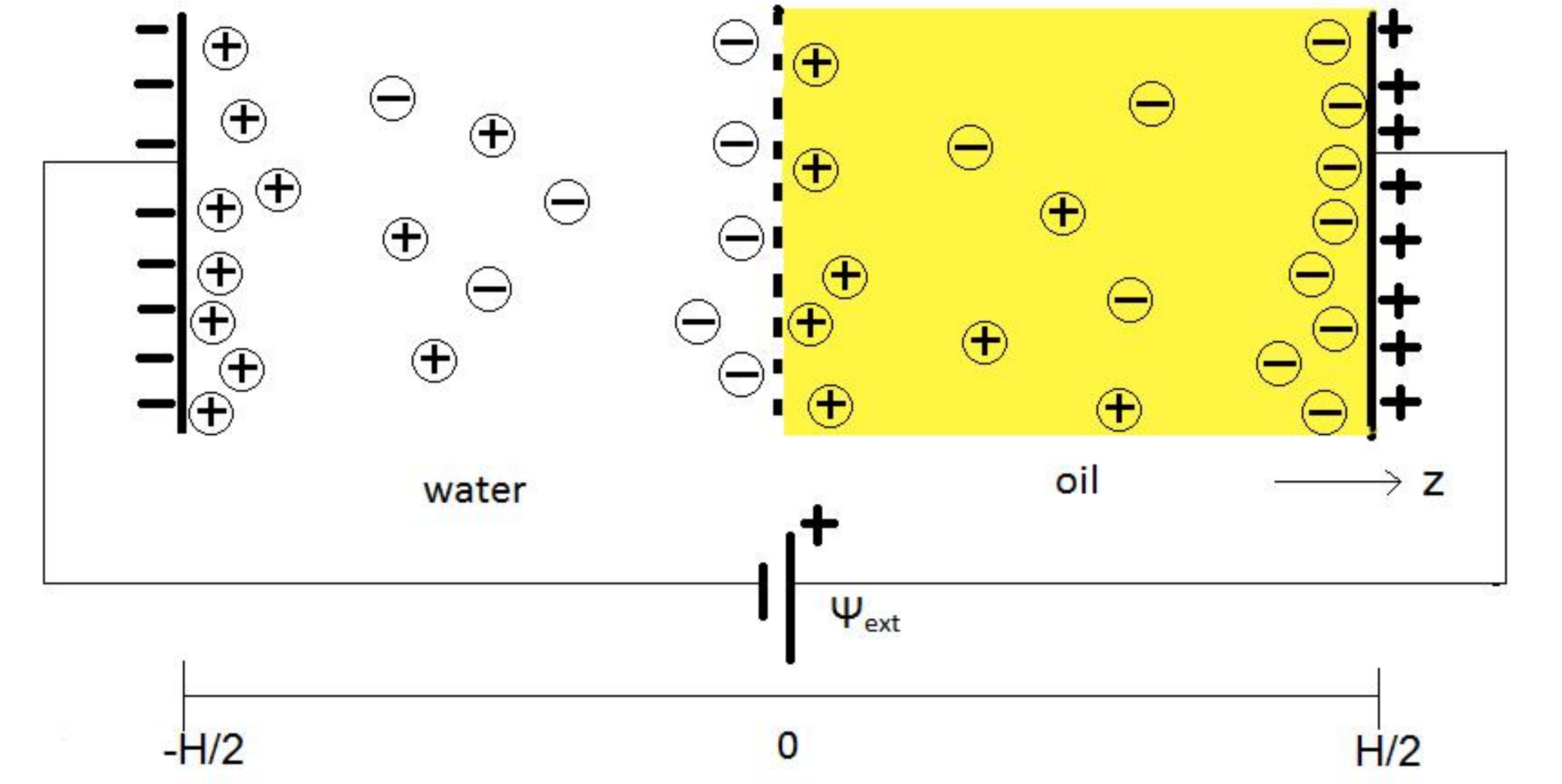}
	\label{fig:potential}
\end{subfigure}
\begin{subfigure}
	\centering
	\includegraphics[width=0.5\textwidth,height=35mm]{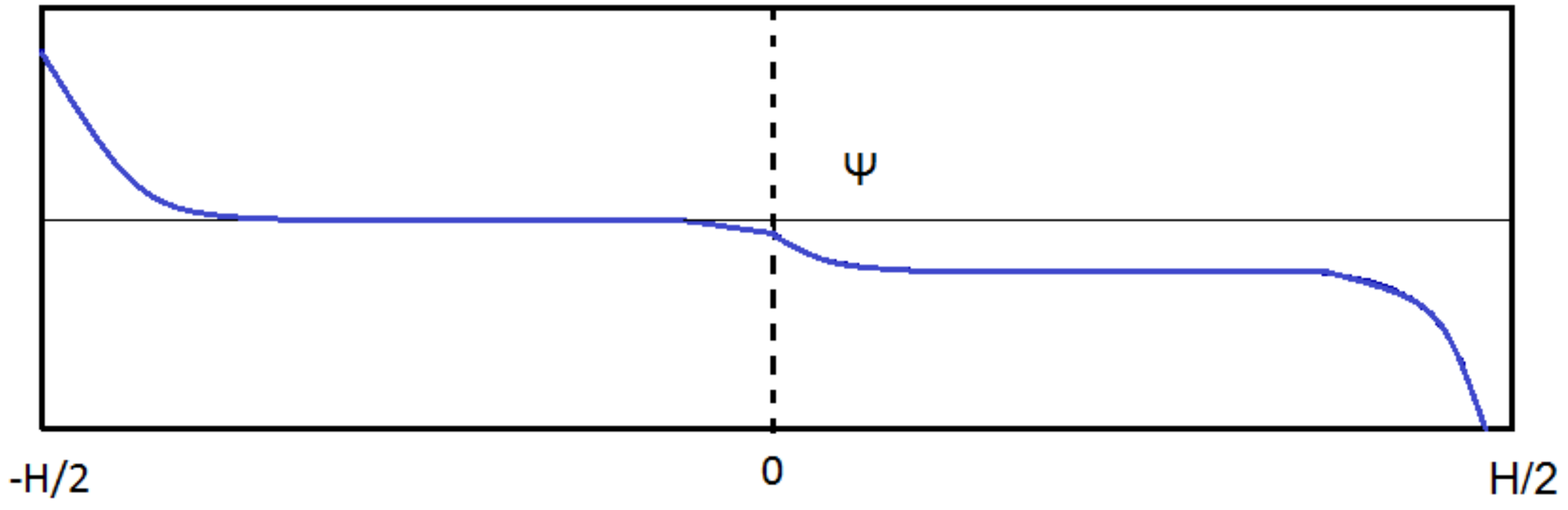}
	\label{fig:potential}
\end{subfigure}
\begin{subfigure}
	\centering
	\includegraphics[width=0.5\textwidth,height=35mm]{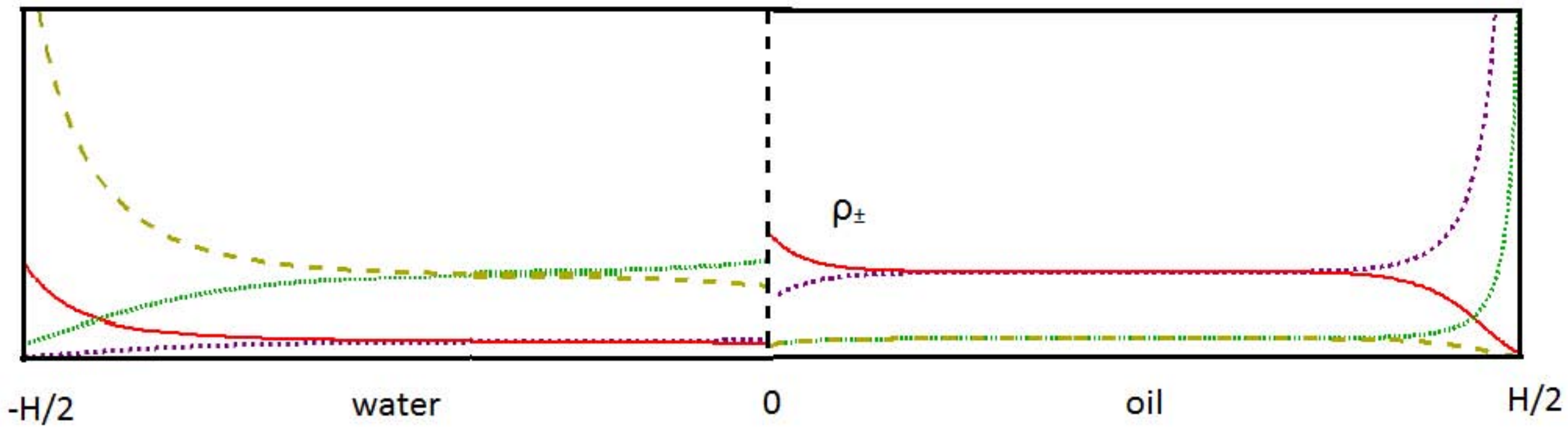}
	\label{densities}
\end{subfigure}
\caption{Model: closed system containing an electrified oil-water interface, and typical electrostatic potential and density profiles for two different salts.}
\label{fig:explanation}
\end{figure}

\subsection{Open electrified oil-water interface}
From the  single planar interface discussed so far we now extend our study and consider an electrolytic cell viewed as {\em three} coupled planar interfaces, as illustrated in Fig.\ref{fig:explanation}. The cell is bounded by two planar electrodes at a distance $H$ from each other in the planes $z=\pm H/2$, and it is filled with two equal volumes of immiscible electrolytes forming an interface in the plane $z=0$.   The two electrolytes will be referred to as ``water'' ($-H/2<z<0$) and ``oil" ($0<z<H/2$), which are both assumed to be 
structureless dielectric liquids fully characterized by their relative dielectric constants $\epsilon_w$ and $\epsilon_o$, respectively, which in turn determine the Bjerrum lengths $\lbw$ in water and $\lbo$ in oil. In order to compare our results with the experiments of Ref.\cite{schloss2012}, we will take $\lbw = 0.72~\mathrm{nm}$ and $\lbo = 5.43~\mathrm{nm}$ throughout this work.  We consider two salts, yielding 4 different ion species which we all assume to be monovalent here. The difference of the ionic solvation free energy between oil and water is for each individual ion species denoted by $k_BTf_{\alpha\pm}$, with $\alpha = \{1,2\}$, such that $f_{\alpha\pm}>0$ for hydrophobic ion species and $f_{\alpha\pm}<0$ for hydrophilic ion species. In line with the point-like nature of the ions and the sharp planar interface between water and oil at $z=0$, we define the external potential for the ion species $\alpha\pm$ as
\begin{equation}\label{eq:selfen}
V_{\alpha\pm}(z) = \begin{cases} 0 &\mbox{if } z < 0 \\
k_BT f_{\alpha\pm}& \mbox{if } z > 0, \end{cases}
\end{equation}
where the zero of solvation free energy is chosen in the water phase. We are interested in the relation between the imposed potential difference $\Delta\Psi$ between the two electrodes and the salt concentration in the electrolytes on the one hand, and the resulting electrode charge densities $+e\sigma$ at $z=-H/2$ and $-e\sigma$ at $z=+H/2$, the ion concentration profiles $\rho_{\alpha\pm}(z)$, and the dimensionless electrostatic potential $\phi(z)$ for $z\in[-H/2,H/2]$ on the other hand. It turns out be convenient, however, to use $\sigma$  as a control variable, and to calculate $\Delta\Psi$.

We will consider macroscopically large cells with two well-defined bulk states, one in the vicinity of $z=-H/4$ in the water phase between the electrode-water interface and the water-oil interface,
and the other in the vicinity of $z=+H/4$ in the oil phase between the water-oil interface and the oil-electrode interface. Asymptotically far from both the electrodes and the water-oil interface, the ion concentration profiles $\rho_{\alpha\pm}(z)$ and the dimensionless electrostatic potential $\phi(z)$ take constant bulk values. It is convenient to gauge the electrostatic potential in bulk water at zero, so $\phi(-H/4)=0$, and to use the ionic bulk concentrations in water, denoted by $\rho_{\alpha\pm}(-H/4)\equiv \rho_{\alpha\pm}^w$,  as control variables. Using the ionic bulk concentrations $\rho_{\alpha\pm}^w$ as control variables implies a grand-canonical treatment of the ions, and in this sense the system is regarded as ``open''. Bulk neutrality imposes that $\sum_{\alpha}(\rho_{\alpha+}^w - \rho_{\alpha-}^w)=0$. With these definitions, the Boltzmann distribution of the ions throughout the cell takes the form 
\begin{equation}
\label{boltzwo}
\rho_{\alpha\pm}(z)=\rho_{\alpha\pm}^w\exp[\mp\phi(z)-V_{\alpha\pm}(z)/k_BT], 
\end{equation}
which in the bulk oil phase leads to bulk ion concentrations $\rho_{\alpha\pm}(H/4)\equiv\rho_{\alpha\pm}^o$ given by
\begin{equation}\label{rhoo}
\rho_{\alpha\pm}^o=\rho_{\alpha\pm}^w\exp[\mp\phi_D-f_{\alpha\pm}].
\end{equation}
Here the so-called Donnan potential of the bulk oil phase, $\phi_D\equiv\phi(H/4)$, follows from the neutrality condition in the bulk oil, $\sum_{\alpha}(\rho_{\alpha+}^o - \rho_{\alpha-}^o)=0$, which can be rewritten as 
\begin{equation}\label{phiD}
\phi_D=\frac{1}{2}\ln\left(\frac{\displaystyle\sum_{\alpha}\rho_{\alpha+}^w\exp(-f_{\alpha+})}{\displaystyle\sum_{\alpha}\rho_{\alpha-}^w\exp(-f_{\alpha-})}     \right).
\end{equation}
Note that Eq.(\ref{phiD}) only holds for monovalent ions. The numerator in the logarithm contains a sum over all cation species and the denominator sums on all anionic species. 

With the (neutral) bulk oil state completely specified in terms of the bulk water state and the self-energy parameters $f_{\alpha}$ in Eqs.(\ref{rhoo}) and (\ref{phiD}), we are now ready to describe the three interfaces. By writing the Poisson equation as
$\phi''(z)=-4\pi\lambda_B^{w,o}\sum_{\alpha}(\rho_{\alpha+}(z) - \rho_{\alpha-}(z))$, where one should take $\lambda_B^w$ for $-H/2<z<0$ and $\lambda_B^o$ for $0<z<H/2$,
and introducing the screening constants 
$\kappa_w^2=4\pi\lambda_B^w\sum_{\alpha}(\rho_{\alpha+}^w+\rho_{\alpha-}^w)$ and  $\kappa_o^2=4\pi\lambda_B^o\sum_{\alpha}(\rho_{\alpha+}^o+\rho_{\alpha-}^o)$, one can write the resulting Poisson-Boltzmann equation as 
\begin{equation}\label{phiint}
\phi''(z) = \begin{cases}\displaystyle \kappa_w^2\sinh\phi(z) &\mbox{if } z < 0 \\
\displaystyle\kappa_o^2\sinh(\phi(z)-\PhiD) & \mbox{if } z > 0, \end{cases}
\end{equation}
with boundary conditions on the interfaces at $z=\pm H/2$ and at $z=0$, and with appropriate asymptotic bulk states at $z=\pm H/4$, given by 
\begin{eqnarray}\label{boundaryint}
\phi'(-H/2)&=&-4\pi\lambda_B^w\sigma; \nonumber\\
\phi(-H/4)&=&0;\nonumber\\
\phi(0^-)&=&\phi(0^+);\nonumber\\
\epsilon_w\phi'(0^-) &=&\epsilon_o\phi'(0^+);\nonumber\\
\phi(H/4)&=&\phi_D;\nonumber\\
\phi'(H/2)&=&-4\pi\lambda_B^o\sigma.
\end{eqnarray}
Here $0^\pm$ is short for the limit to $z=0$ from below $(-)$ or from above $(+)$. 
Typically, $H$ is orders of magnitude larger than either of the two Debye lengths $\kappa_w^{-1}$ and $\kappa_0^{-1}$, such that $H/4$ can be seen as an asymptotic ``infinite" distance from electrodes and/or the oil-water interface. The solution of this set of equations can therefore be written as follows, in analogy with Eq.(\ref{eq:pbsolution}): 
\begin{eqnarray}
\phi(z)=&\nonumber\\
\begin{cases} 
2\log\displaystyle\left\{\frac{1+\gamma_w\exp[-\kappa_w(z+\frac{H}{2})]}{1-\gamma_w\exp[-\kappa_w(z+\frac{H}{2})]}\right\},
&\frac{-H}{2}<z<\frac{-H}{4}; \\
2\log\displaystyle\left[\frac{1+C_w\exp(\kappa_w z)}{1-C_w\exp(\kappa_w z)}\right],
& \frac{-H}{4}<z<0; \\
2\log\displaystyle\left[\frac{1+C_o\exp(-\kappa_o z)}{1-C_o\exp(-\kappa_o z)}\right]
+\PhiD, & 0<z<\frac{H}{4}; \\
2\log\displaystyle\left\{\frac{1+\gamma_o\exp[\kappa_o(z-\frac{H}{2})]}{1-\gamma_o\exp[\kappa_o(z-\frac{H}{2})]}\right\}+\PhiD,
& \frac{H}{4}<z<\frac{H}{2}. \end{cases}\nonumber\\
\label{phiwo}
\end{eqnarray}
The integration constants $\gamma_w$ and $\gamma_o$ are fixed by the boundary conditions at $z=\pm H/2$, and are analogous to the integration constant for the single-electrode case, Eq. (\ref{eq:gamma}), given by 
\begin{eqnarray}\label{gamwo}
\gamma_w &=& \displaystyle\frac{\sqrt{(2\pi\lambda_B^w\kappa_w^{-1}\sigma)^2+1}-1}{2\pi\lambda_B^w\kappa_w^{-1}\sigma}\\
\gamma_o &=&\displaystyle
\frac{\sqrt{(2\pi\lambda_B^o\kappa_o^{-1}(-\sigma))^2+1}-1}
{2\pi\lambda_B^o\kappa_o^{-1}(-\sigma)}.
\end{eqnarray}

The integration constants $C_w$ and $C_o$ follow from the two continuity conditions at $z=0$ (Eq.\ref{boundaryint}), such that
\begin{eqnarray}
\label{Cw0}
C_w &=& \displaystyle\frac{\kappa_o+\exp(\phi_D)\kappa_o+2\exp(\frac{\phi_D}{2})
\kappa_w\frac{\epsilon_w}{\epsilon_o}}{\kappa_o(\exp(\phi_D)-1)}; \nonumber\\
&-&\displaystyle\frac{2\sqrt{k}}{\kappa_o(\exp(\phi_D)-1)}\\
C_o &=& \displaystyle\frac{\kappa_w\frac{\epsilon_w}{\epsilon_o}+\exp(\phi_D)\kappa_w\frac{\epsilon_w}{\epsilon_o}+2\kappa_o\exp(\frac{\phi_D}{2})}{\kappa_w(\exp(\phi_D)-1)}\nonumber\\
&-&\displaystyle\frac{2\sqrt{k}}{\kappa_w(\exp(\phi_D)-1)},
\end{eqnarray}
with
\begin{equation*}\label{eq:kdefenition}
k=\exp(\PhiD)\left(\kappa_o^2+\kappa_w^2\left(\frac{\epsilon_w}{\epsilon_o}\right)^2+2\kappa_o\kappa_w\frac{\epsilon_w}{\epsilon_o}\cosh\left(\frac{\PhiD}{2}\right)\right).
\end{equation*}
The dimensionless charge density $\sigma$ is imposed on the left electrode.\\

We have obtained the closed-form expression for $\phi(z)$ as represented by Eqs.(\ref{phiwo}) in terms of the bulk ion concentrations $\rho_{\alpha\pm}^w$ in the water phase, the energy differences $k_BTf_{\alpha\pm}$, and the electrode charge densities $\pm e\sigma$. The ionic concentration profiles follow explicitly from insertion of $\phi(z)$ into the Boltzmann distribution of Eq.(\ref{boltzwo}). Moreover, two emerging electrostatic quantities can be deduced from our results. The first is the voltage $\Delta\Psi$ between the electrodes, which is given by
\begin{equation}
\label{voltage}
\Delta\Psi = \frac{k_BT}{e}\left(\phi(-H/2)-\phi(H/2)\right).
\end{equation}
The second quantity is the apparent surface charge density $e\sigma'\equiv e\int_{-H/4}^{0}dz\sum_{\alpha}(\rho_{\alpha+}(z) - \rho_{\alpha-}(z))$ at the water-side of the interface, which by global neutrality is the opposite of the apparent surface charge density at the oil-side of the interface. Using the Poisson equation and applying the Gauss law, we find
\begin{equation}
\label{sigma2}
\sigma'=\frac{\phi'(0^-)}{4\pi\lambda_B^w}\hspace{1cm}\left(=\frac{\phi'(0^+)}{4\pi\lambda_B^o}\right).
\end{equation}
Moreover, for later reference we will also calculate the adsorption $\Gamma_{\alpha\pm}^{ab}$ of cation/ion species $\alpha$ to the $a-b$ interface, where $a-b$ can refer to the electrode-water (e-w), the water-oil (w-o), or the oil-electrode (o-e) interface. In line with Eq.(\ref{eq:Gamma}) we now find
\begin{eqnarray}
\Gamma_{\alpha\pm}^{ew}&=&\frac{\mp4\rho_{\alpha\pm}^w}{\kappa_w}\frac{\gamma_w(\sigma)}{1\pm\gamma_w(\sigma)};\nonumber\\
\Gamma_{\alpha\pm}^{ow}&=&\frac{\mp4 \rho_{\alpha\pm}^w}{\kappa_w}\frac{\gamma_w(\sigma')}{1\pm\gamma_w(\sigma')}+
\frac{\mp4 \rho_{\alpha\pm}^o}{\kappa_o}\frac{\gamma_o(\sigma')}{1\pm\gamma_o(\sigma')};\nonumber\\
\Gamma_{\alpha\pm}^{oe}&=&\frac{\mp4 \rho_{\alpha\pm}^o}{\kappa_o}\frac{\gamma_o(-\sigma)}{1\pm\gamma_o(-\sigma)},
\end{eqnarray}
which with Eqs.(\ref{gamwo}) gives analytic expressions in terms of the control variables.

For fixed ion concentrations in bulk water, and for fixed surface charge density on the electrodes $\pm e\sigma$ we have thus found explicit results for the voltage $\Delta\Psi$ between the electrodes, the Donnan potential $\Psi_D=(k_BT/e)\phi_D$ between water and oil, the ion concentrations $\rho_{\alpha\pm}^o$ in the bulk oil phase, the degree of charge separation $\sigma'$ at the oil-water interface and the ion adsorption at the three interfaces of the cell. Note that for convenience we use $\sigma$ as a control variable with a resulting voltage $\Delta\Psi$ that we calculate, although we could have reversed this by fixing $\Delta\Psi$ and calculating the resulting electrode charge $\sigma$, a procedure that would be closer to the experimental reality. However, the one-to-one relation between voltage and charge renders both alternatives equivalent. 

\subsection{Closed electrified water-oil interface}
Interestingly,  for fixed $f_{\alpha\pm}$ our analysis of the open system above also reveals that the Donnan potential  $\phi_D$ as defined in Eq.(\ref{phiD}) {\em only} depends on the set of concentrations $\rho_{\alpha\pm}^w$ in bulk water, and {\em not} on the cell size $H$ or the electrode charge density $\sigma$. The same holds for the ion concentrations $\rho_{\alpha\pm}^o$ in bulk oil given in Eq.(\ref{rhoo}), for the integration constants $C_w$ and $C_o$ in Eq.(\ref{Cw0}), and hence also for the interfacial surface charge density $\sigma'$. In other words, for fixed $\rho_{\alpha\pm}^w$ all thermodynamic and electrostatic properties of the ``electrified'' oil-water interface are independent of the electrode separation, charge, and voltage. This independence is easy to understand qualitatively if one realizes that the electrode charge is completely screened beyond a few Debye lengths $\kappa_w^{-1}$ and $\kappa_o^{-1}$, which are assumed to be much smaller than the cell size $H$. However, in the experiments of Ref.\cite{schloss2012} it is argued that the properties of a water-oil interface can be tuned significantly by applying a voltage, even if $H$ is in the centimetre regime and the Debye lengths in the nanometer regime, i.e. in a regime where the assumption of asymptotically large $H$ should be perfectly valid. The present theory can only explain the tunable electrified water-oil interface if the electrode charging process affects the bulk ion concentrations. This seems unlikely at first sight, in view of the macroscopic (cm-range) size of the cell. However, below we will show that charging the electrodes while treating the ions either canonically or grand-canonically makes a qualitative difference. In fact we will identify a new length scale $H^*$, of the order of $\sigma\exp(|f|)/\rho^w$, which can be of order mm-m for typical self energies $|f|=\mbox{min}_{\alpha\pm}\{|f_{\alpha\pm}|\}\simeq10-20$, typical electrode charges $\sigma\simeq\mbox{nm}^{-2}$, and typical salt concentrations $\rho_{\alpha\pm}^w\simeq\mbox{mM-M}$. Only for $H\gg H^*$ is the system size large enough for the charging process of the electrodes to be viewed as grand canonical in the ions. For smaller cells a canonical treatment turns out to be appropriate. For that reason we now consider a closed system with fixed numbers $N_{\alpha\pm}$ of cations/anions of species $\alpha$.

Denoting the total surface area of an electrode by $A$, such that the volume of the cell is $AH$, we can write
\begin{equation}\label{eq:N}
N_{\alpha\pm}=\frac{AH}{2}\rho_{\alpha\pm}^w + \frac{AH}{2}\rho_{\alpha\pm}^0 + A\left(\Gamma_{\alpha\pm}^{ew}+\Gamma_{\alpha\pm}^{ow}+\Gamma_{\alpha\pm}^{oe}\right),
\end{equation}
where we note that $\rho_{\alpha\pm}^o$ and all $\Gamma_{\alpha\pm}^{ab}$ are explicitly known in terms of the set of bulk water concentrations $\{\rho_{\alpha\pm}^w\}$. In other words,  the right hand side of Eq.(\ref{eq:N}) is an explicit function of these variables, and hence we can view Eq.(\ref{eq:N}) as a closed set of equations to calculate $\rho_{\alpha\pm}^w$ for given 
$N_{\alpha\pm}$ at fixed $A$ and $H$. 

We will focus on the specific case of an inorganic (hydrophilic)  salt (e.g. NaCl) and an organic (hydrophobic) salt, corresponding to $\alpha=1$ and $\alpha=2$ respectively. We assume complete dissociation and therefore set $N_{1+}=N_{1-}=AH\rho_1$ and $N_{2+}=N_{2-}=AH\rho_2$, where $\rho_1$ and $\rho_2$ are the
(imposed) overall concentration of the inorganic and organic salt in the cell, respectively. By inserting these definitions into Eq.(\ref{eq:N}) the dependence on the surface area $A$ cancels, and we can apply standard numerical root finding procedures to calculate the four unknown bulk water concentrations  $\rho_{\alpha\pm}^w$ for fixed $\rho_1$, $\rho_2$, $\sigma$, $H$, and $f_{\alpha\pm}$.
\begin{figure}[h!]
\centering
\includegraphics[width=0.5\textwidth]{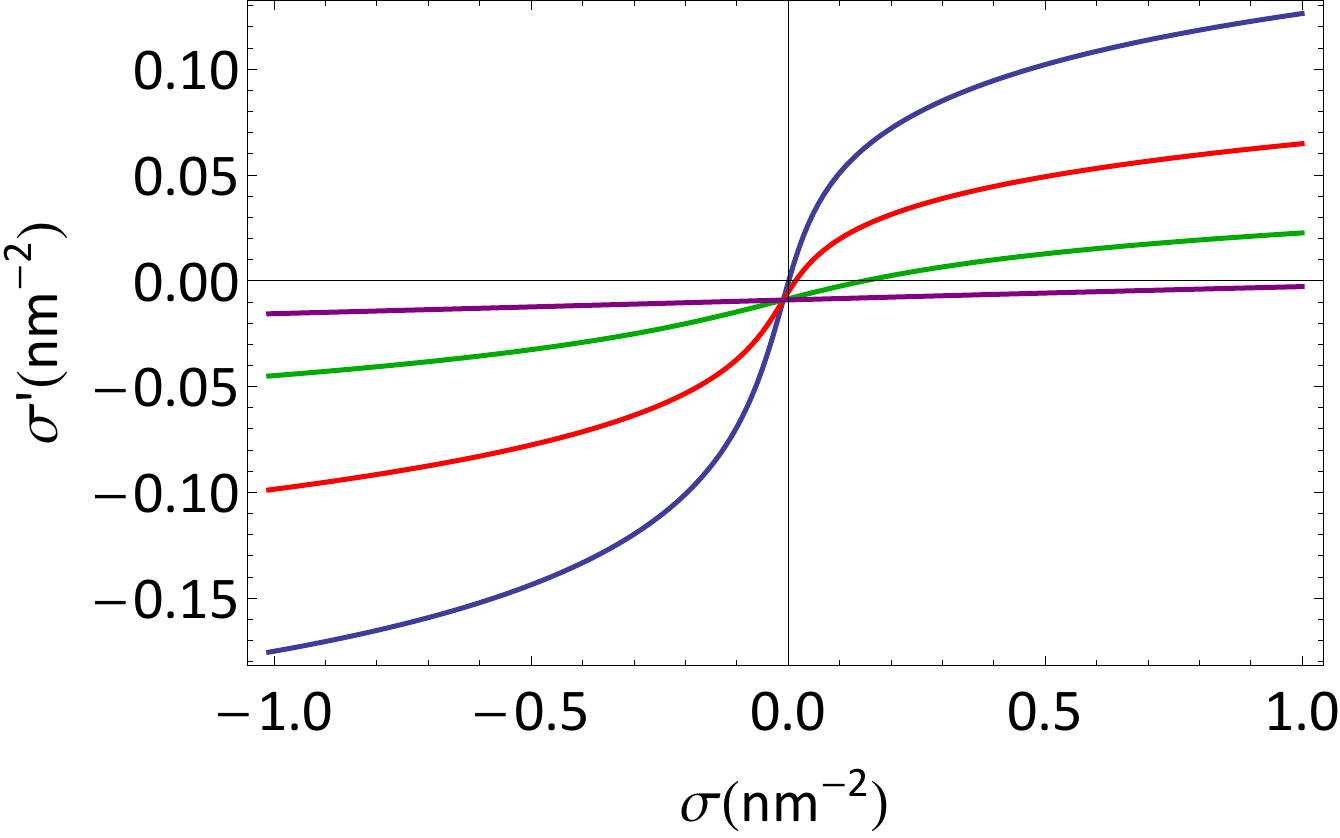}
\caption{The charge accumulation $\sigma'$ at the interface, as defined in Eq. (\ref{sigma2}) as a function of the electrode surface charge density $\sigma$ for ionic self-energies $f_{\alpha\pm}=(-22,-12,12,18)$, for canonical inorganic and organic ion concentrations $\rho_\alpha=10~\mathrm{mM}$. The flat curve corresponds to $H=10^6~\mathrm{nm}$; increasingly steeper graphs show $H=10^5$, $H=10^4$ and $H=10^3~\mathrm{nm}$, respectively.}
\label{fig:systemsize1}
\end{figure}
\section{Numerical results}
The relation between $\sigma'$ and $\sigma$ is useful in understanding the response of the system to charging the electrodes. We will distinguish between two limiting cases, referred to as the canonical and the grand-canonical limit. The grand-canonical limit is attained when the oil-water interface is effectively decoupled from the electrodes by a sufficiently large distance $\frac{H}{2}$: $\sigma'$ is nonzero but does not depend on $\sigma$. By contrast, $\sigma'$ can be manipulated one-to-one by $\sigma$ in the canonical limit. For fixed self-energies $f_{\alpha\pm}$ and canonical bulk densities $\rho_\alpha=\frac{N_\alpha}{A H}$ we study the effects of the system size, characterized by the separation $H$ between the two planar electrodes. Fig.\ref{fig:systemsize1} shows the relation between the oil-water charge $\sigma'$ and $\sigma$ for $H=10^3, 10^4, 10^5$ and $10^6~\mathrm{nm}$, for system parameters $f_{\alpha\pm}=(-22,-12,12,18)$ and $\rho_1 = \rho_2=10~\mathrm{mM}$. The largest value of $H$ clearly shows a relatively low interfacial charge $\sigma'$, that is, moreover, only weakly dependent on the electric charge.

\begin{figure}[!ht]
\centering
\includegraphics[width=0.5\textwidth]{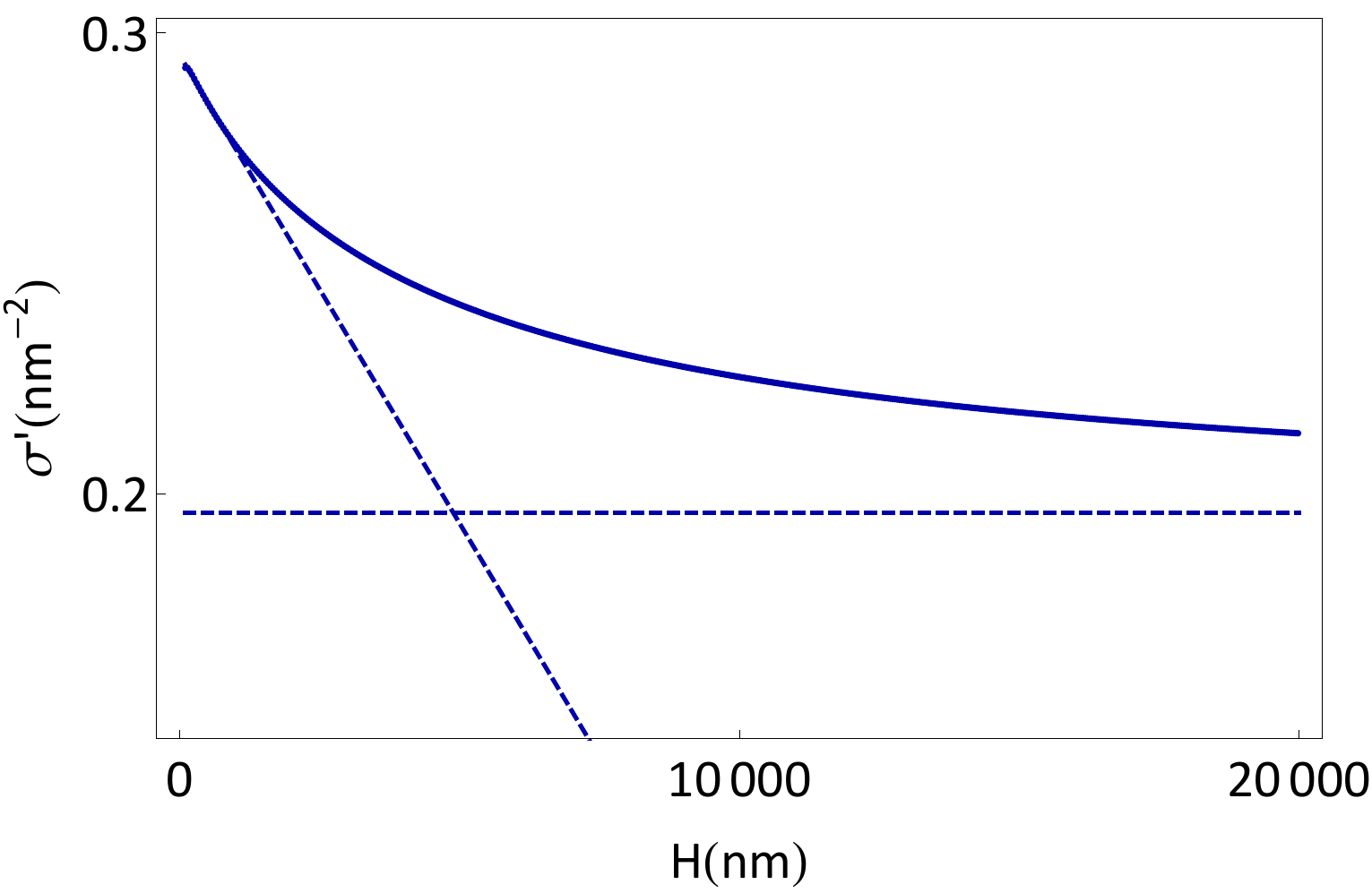}
\caption{The oil-water charge density $\sigma'$ as a function of the electrode-electrode separation $H$ (solid line) and its value in the limit of infinite $H$ (dashed) for $\rho_1 = \rho_2 =10~\mathrm{mM}$, $\sigma=0.3~\mathrm{nm^{-2}}$ and $f_{\alpha\pm}=(-15,0,0,15)$. The crossover length $H^{*}$ is given by the value of $H$ where the line tangent to the curve at small $H$ and the limiting (grand-canonical) value of $\sigma'$ intersect.}
\label{fig:excrossover}
\end{figure}

For fixed values of ionic self-energy differences $f_{\alpha\pm}$ the crossover length $H^{*}$ marks the transition between the canonical and the grand-canonical regime. We will determine the latter as the point where the tangent line to $\sigma'$ ($H$) for small ($\simeq 100~\mathrm{nm}$) values of $H$ equals the grand-canonical value of $\sigma'$, as is illustrated in Fig. \ref{fig:excrossover}.

\begin{figure}[!ht]
\centering
\includegraphics[width=0.5\textwidth]{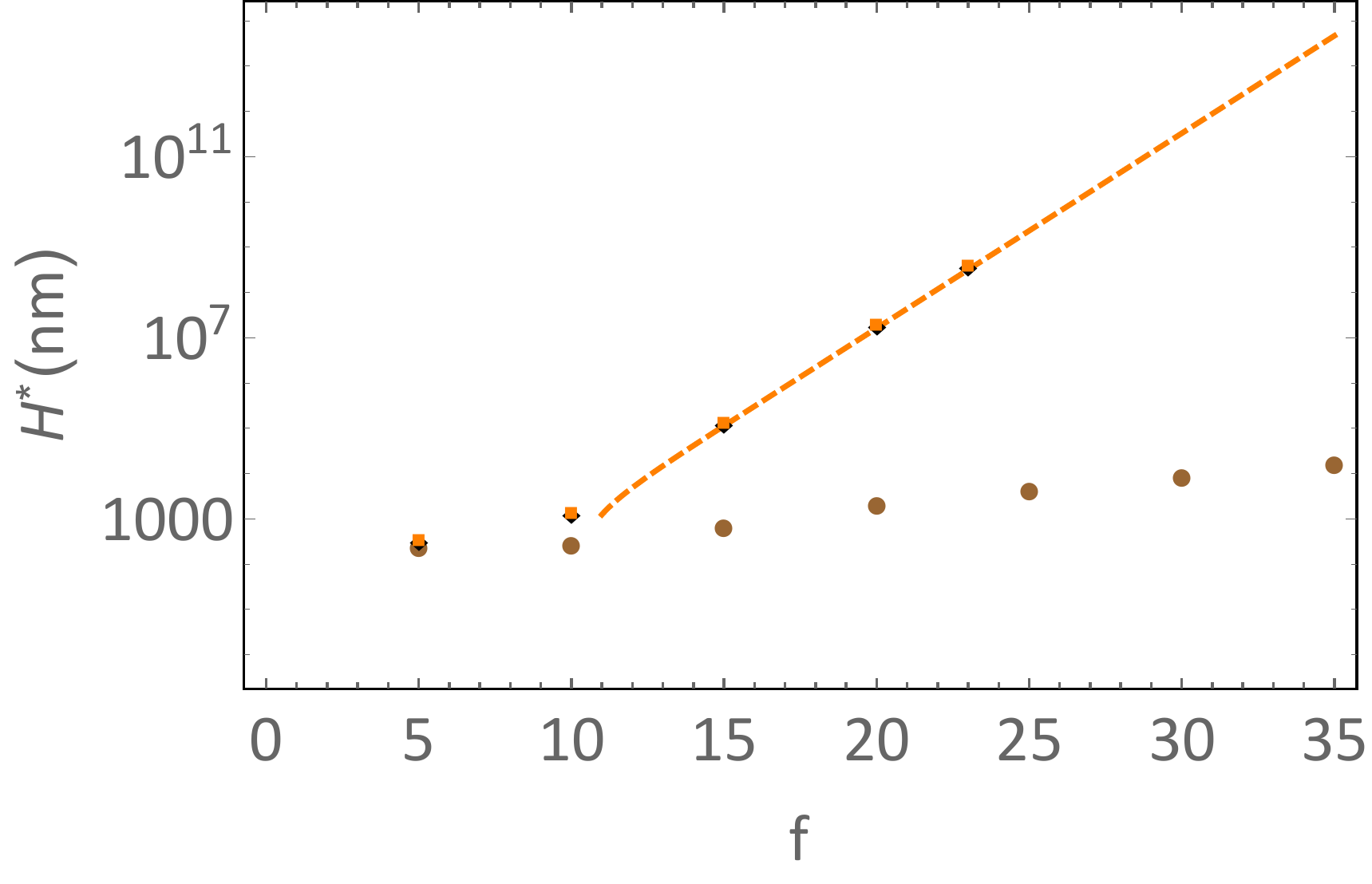}
\caption{Grand-canonical to canonical crossover electrode-electrode separation $H^*$ as a function of the self-energy parameter $f$, for electrode charge density $\sigma=0.3~\mathrm{nm^{-2}}$ and $\rho_1=\rho_2=1~\mathrm{M}$. The upper curve signifies the set of self-energies $(-f,-f,f,f)$ and the two (almost indistinguisable) lower curves represent the two sets of self-energies $(-f,0,f,f)$ and $(-f,0,0,f)$, which contain at least one ionic species without a preference for either phase. An exponential fit was made for the upper curve, for the range $|f|\geq 11$.}
\label{fig:crossover2}
\end{figure}

Crossover lengths for $f_{\alpha\pm}$ of the forms $(f_{1-},f_{1+},f_{2-},f_{2+}) =$  $(-f,0,0,f)$, $(-f,0,f,f)$ and $(-f,-f,f,f)$ are shown in Fig. \ref{fig:crossover2}, suggesting that $H^*\simeq a\exp f-c$ in the latter case. This can be understood as follows: if all ion species have a strong preference for their native phase then the majority of the ions will remain in that phase, e.g. $\rho_{1\pm}^w \approx 2 \rho_1$ and $\rho_{2+}^o \approx 2 \rho_2$. The relatively tiny fraction that migrates from the native phase to the other phase is, following from Equation 7, described by $\rho_{1\pm}^o \approx 2 \rho_1 \exp[\mp \phi_D - f_{1\pm}]$ and $\rho_{2\pm}^w \approx 2 \rho_2 \exp[\pm \phi_D + f_{2\pm}]$. As we have seen before, the Donnan potential that enters here implicitly depends on these ion densities. However, as we are interested in the dependence of $\sigma'$ on $H$ for sufficiently small $H$, we may approximate $\phi_D$ from the assumption $\sigma'\approx\sigma$. Equation 2 can be applied to find the electrostatic potential difference between the bulk water phase charged interface, as well as the electrostatic potential difference between the charged interface and the bulk oil phase. The sum of these contributions adds up to the Donnan potential, which for the parameters in Figure 4 is $\phi_D \approx -2.6$.  Charge neutrality dictates that the apparent charge of the interface is related to the amount of ions that have migrated, $\sigma' \approx \sigma - \frac{H}{2}\left(\rho_{1+}^o + \rho_{1-}^o + \rho_{2+}^w - \rho_{2-}^w\right)$. This approximation holds for the case that the double layers occupy only a small portion of both phases, which is increasingly accurate for $H\gg\kappa_o^{-1}$ and $H\gg\kappa_w^{-1}$. Although all four ion species migrate, they do so in different proportions. Those that experience a low self-energy penalty and/or those that move down in electrostatic energy can be the dominant migrating species and therefore solely determine $\sigma'$ for small $H$. For the special case (--$f$,--$f$,$f$,$f$) that we consider in Figure 4 the process is governed by the cations that migrate from the water phase to the oil phase as well as the anions that migrate from the oil phase to the water phase. We therefore find $\sigma'(H) \approx \sigma - H (\rho_1 + \rho_2) \exp[f- |\phi_D| ]$, and thus
\begin{equation}
H^* \approx \frac{\sigma  \exp[f-|\phi_D|]}{(\rho_1 + \rho_2) }.
\end{equation}
The dashed line in Figure 4 represents this analytical approximation to $H^*$, demonstrating very good agreement with our numerical approach. We also include numerical data corresponding to the parameters sets $(-f,0,f,f)$ and $(-f,0,0,f)$ in the Figure, which turn out to be barely distinguishable from each other. Our results therefore indicate that the presence of one ion species without a preference for water or oil will affect the system in much the same way as two species with this zero self-energy difference, resulting in a decrease of the crossover length by orders of magnitude in both cases. The analytical approach that we described above cannot be applied to quantitatively estimate $H^*$ for these cases, since some of the ion species have no preference for either phase. Nevertheless, it can be understood from e.g. Equation 20 that decreasing $f$ to small values yields a dramatic decrease in $H^*$, which is in line with the observations.\\

The experiments of Ref.\cite{schloss2012} formed a direct motivation to study the electrolytic cell in more detail. An electrolytic cell of length $H=4~\mathrm{cm}$ containing aqueous ($\epsilon_w=78.54$) and organic ($\epsilon_o=10.43$) electrolyte solutions is considered at $T=294~\mathrm{K}$. Sodium chloride was dissolved in water to produce a 10 mM solution. A solution of BTPPATPFB in DCE was prepared at a concentration of 5 mM. Because of the low dielectric constant of DCE only partial dissociation into $\mathrm{BTPPA^{+}}$ and $\mathrm{TPFB^{-}}$ occurs, producing an organic solution with a dissociated ionic concentration of 2.7 mM \cite{schloss2012}.
The differences between the bulk values of the potentials of mean force (PMFs), which were modeled by molecular dynamics (MD) simulations, are given by \cite{schlossprivate}
\begin{align*}
(f_{\mathrm{TPFB^{-}}},f_{\mathrm{BTPPA^{+}}},f_{\mathrm{Cl^{-}}},f_{\mathrm{Na^{+}}})\nonumber\\
=(-29.9,-22.9,22.3,21.2).
\end{align*}
In Figure 5 we examine three parameter sets of the self-energies
\begin{enumerate}
\item\label{enself2} $f_{\alpha\pm}=(-29.9,-22.9,22.3,21.2)$ (experimental values of the Gibbs free energies)
\item\label{enself1} $f_{\alpha\pm}=(-29.9,0,0,21.2)$ (Gibbs free energies of transfer, where self-energies of $\mathrm{BTPPA^{+}}$ and $\mathrm{Cl^{-}}$ have been set to zero)
\item\label{enself3} $f_{\alpha\pm}=(-49.9,-42.9,42.3,41.2)$ (Gibbs free energies after addition of $20$: the canonical limit is appropriate for these self-energies).
\end{enumerate}
Fig. 5 shows that the set of experimental self-energy parameters (set 1) gives rise to an oil-water interfacial charge density $\sigma'$ that can indeed be tuned throughout the interval $[-0.15,+0.12]$ nm$^{-2}$ by the electrode charge $\sigma\in[-0.3,+0.3]$nm$^{-2}$, very strongly so in the small-$\sigma$ regime $|\sigma|<0.05$ nm$^{-2}$ where $\sigma'=\sigma$, and only weakly for larger $\sigma$ where $\sigma'$ approaches a saturation regime. This tenability of parameter set 1 is to be contrasted by the behavior of set 2 with two ion species having a vanishingly small self-energy, which gives rise to a large interfacial charge density $\sigma'=+0.33$nm$^{-2}$ that is, however, {\em not at all } tunable by the electrode charge $\sigma$. For set 3, with its additional $20 k_BT$ of self-energy for all  ionic species (which prevents essentially any ion migration to the unfavored solvent), we see from Fig.5 that perfect tuning is possible with $\sigma'=\sigma$ in the whole  regime of $\sigma$ that is considered.\\
The grand-canonical behavior that is revealed by the self-energies of set 2 (for the present ion concentration and stem size) in Fig.5 is also observed for any self-energy set that contains at least one vanishing self-energy, since in such a case the presence/absence of this 'transferable' ion can take care of the screening of the electrodes, thereby decoupling the oil-water interface from the electrodes. We also find that a minimum value of about $|f_{\alpha\pm}|>20$ is needed for all ion species in order to be able manipulate $\sigma'$ by $\sigma$ to a degree comparable that of set 1.
\begin{figure}[!ht]
\centering
\includegraphics[width=0.5\textwidth]{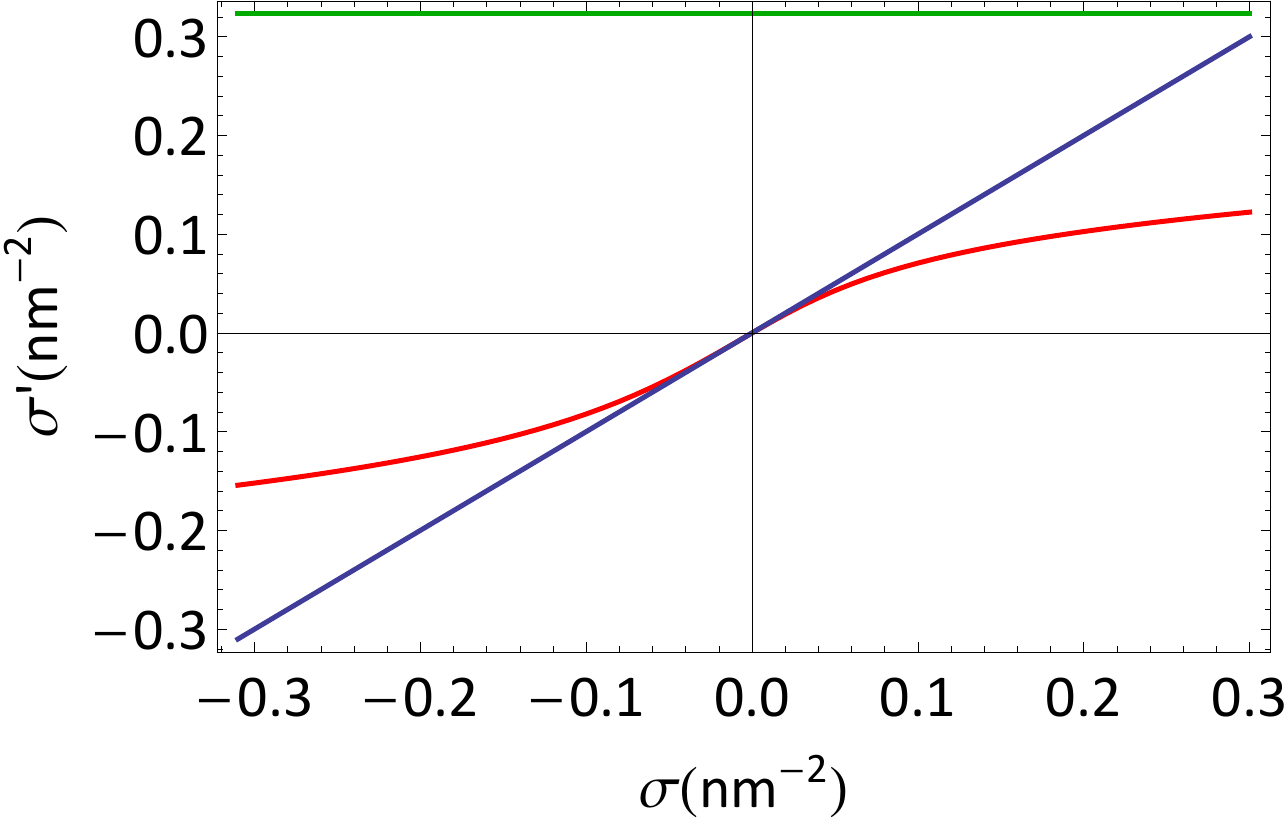}
\caption{Oil-water interface charge density $\sigma'$ as a function of the electrode charge density $\sigma$ for electrode separation $H=4~\mathrm{cm}$, contrasting $f_{\alpha\pm}=(-29.9,0,0,21.2)$ (constant: grand-canonical), $f_{\alpha\pm}=(-29.9,-22.9,22.3,21.2)$ (almost entirely canonical) and $f_{\alpha\pm}=(-49.9,-42.9,42.3,41.2)$ (linear: canonical). Canonical densities are $\rho_1=2.7~\mathrm{mM}$ (oily solution) and $\rho_2=5~\mathrm{mM}$ (aqueous solution).}
\label{fig:TICsigma2}
\end{figure}

\section{Surface tension}
The interfacial tension $\gamma^{int}$ as obtained from density functional theory reads
\begin{align}\label{eq:DFT}
\gamma^{int}/k_B T &= \rho^w \int_{-H/4}^{0}\mathrm{dz}~\left[\phi(z)\sinh(\phi(z)) - 2\cosh(\phi(z))+2\right]\nonumber\\
&+ \rho^o \int_{0}^{H/4}\mathrm{dz}~\left[\tilde\phi(z)\sinh(\tilde\phi(z)) - 2\cosh(\tilde\phi(z))+2\right]\nonumber\\
&+\frac{1}{2} \sigma' \phi_{\mathrm{D}},
\end{align}
where $2\rho^{o/w} = \rho_{1+}^{o/w} + \rho_{1-}^{o/w} + \rho_{2+}^{o/w} + \rho_{2-}^{o/w}.$ A derivation of Eq. (\ref{eq:DFT}) can be found in the appendix.\\
 
\begin{figure}[ht!]
\centering
\includegraphics[width=0.5\textwidth]{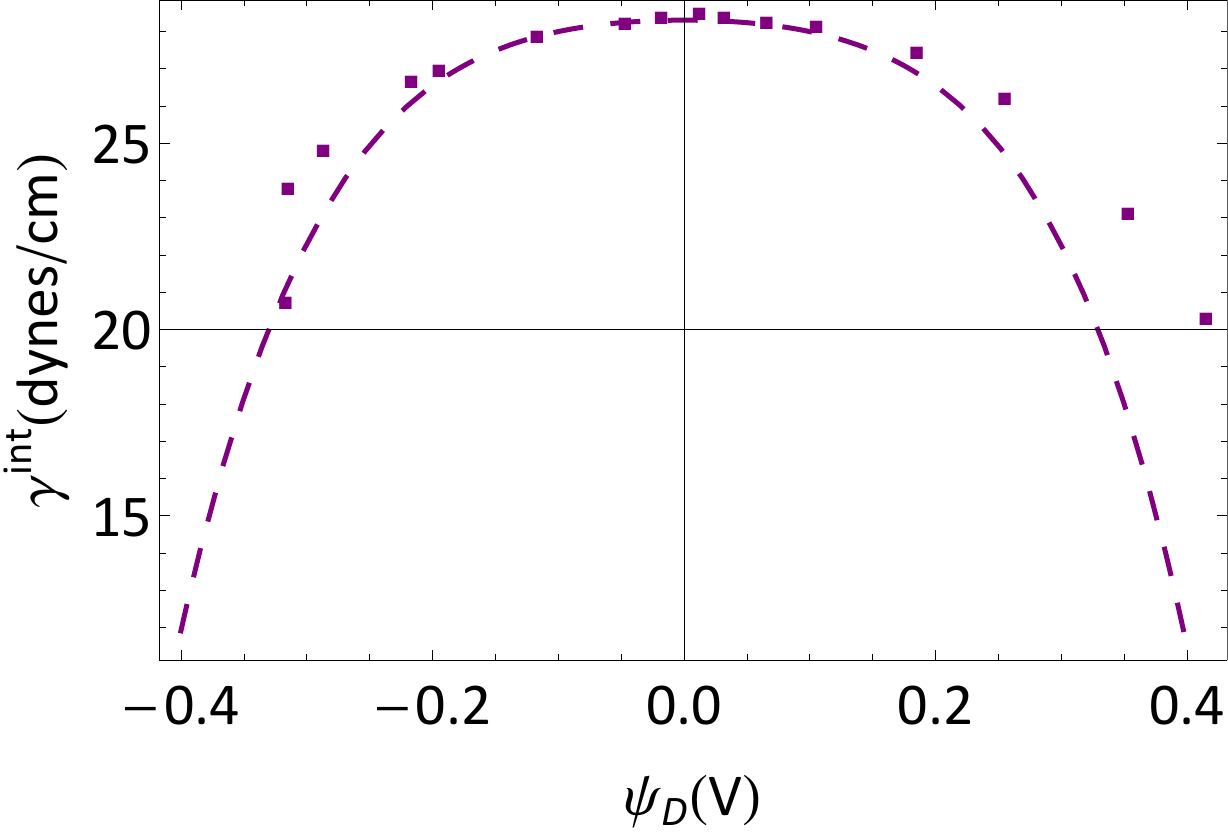}
\caption{The interfacial tension $\gamma^{int}$ as a function of the Donnan potential as derived from density functional theory (dotted line) and as measured at the oil-water interface of an electrolytic cell \cite{schloss2012supp}, for $f_{\alpha\pm}=(-29.9,-22.9,22.3,21.2)$, $\rho_1=2.7~\mathrm{mM}$, $\rho_2=10~\mathrm{mM}$ and electrode-electrode separation $H=4~\mathrm{cm}$. The potential $\phi_{\mathrm{D}}$ is shifted by the potential of zero charge $\Delta\Phi^{pzc}$ \cite{schloss2012supp}. An experimental constant of 28.3 dynes/cm was added to the results of the DFT in order to account for the bare oil-water tension.}
\label{fig:Lippmanndata}
\end{figure}

Surface tension measurements and the interfacial tension as derived from density functional theory (DFT) are compared in Fig. \ref{fig:Lippmanndata}. Qualitatively similar results are found for small values of the potential difference $\PhiD$ between the bulk phases.

\section{Conclusion}
We have applied nonlinear Poisson-Boltzmann theory to a liquid-liquid interface within a closed system bounded by two electrodes with an adjustable electric potential. We have considered the solution of four ion species with different solvation free energies (self energies) and calculated the equilibrium distribution of the salts. We have studied the electric charge accumulation at the liquid-liquid interface. We found that two factors determine the tunability of the interfacial layer: the solvation energies and the separation between the electrodes. For small self energies and a macroscopic electrode-electrode distance, the external potential leaves the interface unaffected, whereas complete control over the surface can be gained for the conjugate combination. We have defined a crossover length $H^{*}$, that marks the transition between these extremes as a function of the solvation energies. We conjecture that an exponential relation exists between the distance $H^{*}$ and the self energies. The presumed dependence of $H^{*}$ on the electrode charge $\sigma$ and the concentrations of the salts $\rho_{1,2}$ merits further investigation. We have derived an expression for the surface tension at the liquid-liquid interface for our model, which we found to be in reasonably good agreement with experimental data obtained by Laanait et al. \cite{schloss2012}.

\appendix
\section{Derivation of Eq. \ref{eq:DFT}}
We consider an open, non-electrified Coulombic system containing an organic and an inorganic salt, made of inorganic ions ($\alpha=1$) and organic ions ($\alpha=2$). If we assume two bulk phases and an interface at $z=0$, the grand potential for this system may be written as
\begin{eqnarray}\label{eq:appgpf}
\Omega[\{\rho_{\alpha\pm}\}]/k_B T=\sum_{\alpha\pm}A\int_{-H/2}^{H/2} \mathrm{dz}~\rho_{\alpha\pm}(z)\nonumber\\
\left[\ln\left(\frac{\rho_{\alpha\pm}(z)}{\rho_{\alpha\pm}^w}\right)-1+\frac{V_{\alpha\pm}}{k_B T}\pm\frac{1}{2}\phi(z,[\{\rho_{\alpha\pm}])\right],
\end{eqnarray}
where $A$ denotes the surface area of the interface and $V_{\alpha}$ is defined as in Eq. (\ref{eq:selfen}). We are primarily interested in the densities $\rho_{\alpha}$, which are sensitive to the presence of electrodes. Functional differentiation of Eq. (\ref{eq:appgpf}) with respect to $\rho_{\alpha}$ yields the Boltzmann distributions Eq. (\ref{boltzwo}).
The relative potential and bulk concentration in oil are respectively defined as:
\begin{eqnarray}\label{eq:appdefs}
\tilde{\phi}(z) &=&\phi(z) - \phi_{\mathrm{D}}\nonumber\\
\rho_{\alpha\pm}^o&=&\rho_{\alpha\pm}^w\exp(\mp\phi_{\mathrm{D}}-f_{\alpha\pm}),
\end{eqnarray}
where $\phi_D$ is the Donnan potential as given by Eq.(\ref{phiD}). We define the grand canonical potential for the interface as the contribution to expression (\ref{eq:appgpf}) bounded by $-\frac{H}{4}$ and $\frac{H}{4}$. We distinguish between the water ($z<0$) and oil ($z>0$) phases, such that $\Omega/k_B T=\Omega_w/k_B T+\Omega_o/ k_B T$. Substitution of equations (\ref{boltzwo}) and (\ref{eq:appdefs}) into equation (\ref{eq:appgpf}) yields
\begin{align}
&\Omega_{w}/ k_B T = A \int_{-H/4}^{0}\mathrm{dz}~\left[\sum_{\alpha\pm} \rho_{\alpha\pm}^w \exp (\mp\phi(z))\left(\mp\frac{\phi(z)}{2} -1\right)\right];\\
&\Omega_{o}/ k_B T\nonumber\\
&= A \int_{0}^{H/4}\mathrm{dz}~\left[\sum_{\alpha\pm}\rho_{\alpha\pm}^w \exp (\mp\phi(z) - f_{\alpha\pm})\left(\mp\frac{\phi(z)}{2} -1\right)\right]\nonumber\\
&= A \int_0^{H/4}\mathrm{dz}~\left[\sum_{\alpha\pm} \rho_{\alpha\pm}^o \exp (\mp\tilde\phi(z))\left(\mp\frac{(\tilde \phi(z) +\phi_{\mathrm{D}}) }{2} -1 \right)\right].
\end{align}
Hence the Donnan potential makes a nontrivial contribution to the grand canonical potential (\ref{eq:appgpf}) in the interfacial region. Whilst in the water phase we find
\begin{equation}
\displaystyle
\Omega_w/ k_B T = A \rho^w \int_{-H/4}^{0}\mathrm{dz}~\left[\phi(z)\sinh(\phi(z)) - 2\cosh(\phi(z))\right],\label{eq:Omegaw} 
\end{equation}
in the oil phase we obtain
\begin{align}
\Omega_o /k_B T &= A \rho^o \int_{0}^{H/4}\mathrm{dz}~\left[\tilde\phi(z)\sinh(\tilde\phi(z)) - 2\cosh(\tilde\phi(z))\right]\nonumber\\ 
&-\frac{1}{2} A \int_0^{H/4}\mathrm{dz}~\left[\sum_{\alpha} (\rho_{\alpha+}(z) - \rho_{\alpha-}(z)) \phi_{\mathrm{D}}\right]\label{eq:Omegao} ,
\end{align}
where $2\rho^{o/w} = \rho_{1+}^{o/w} + \rho_{1-}^{o/w} + \rho_{2+}^{o/w} + \rho_{2-}^{o/w}$. Note that the integrand in Eq.~(\cite{eq:Omegaw}) reduces to $-2$ when the electrostatic potential reaches its vanishing bulk value. This also holds for the first integrand in Eq.~\cite{label:eq:Omegao} upon the electrostatic potential reaching the Donnan potential. By subtracting these bulk contributions we can identify the interfacial energy density,
\begin{align*}
\gamma^{int}/k_B T &= \rho^w \int_{-H/4}^{0}\mathrm{dz}~\left[\phi(z)\sinh(\phi(z)) - 2\cosh(\phi(z))+2\right]\\
&+ \rho^o  \int_{0}^{H/4}\mathrm{dz}~\left[\tilde\phi(z)\sinh(\tilde\phi(z)) - 2\cosh(\tilde\phi(z))+2\right]\\
&+ \frac{1}{2}\sigma' \phi_{\mathrm{D}},
\end{align*}
for which we have used that the charge density in the double layer on the oil side of the interface will exactly balance the charge density of the interface,  $\int_0^{H/4}\mathrm{dz}~\left[\sum_{\alpha} (\rho_{\alpha+}(z) - \rho_{\alpha-}(z)) \right] = -\sigma'$.

\end{document}